\documentclass[aps,prb,superscriptaddress,twocolumn,showpacs]{revtex4}
\usepackage{amsmath,amssymb,mathrsfs}
\usepackage[dvips]{graphicx}

\newcommand{\anticomm}[2]{\left\{#1,#2\right\}}
\newcommand{\ket}[1]{\left|#1\right\rangle}
\newcommand{\bra}[1]{\left\langle#1\right|}

\newcommand{\ii}{\text{i}}

\allowdisplaybreaks[4]

\begin{document}
\title{Local density of states of a quarter-filled one-dimensional\\ Mott
insulator with a boundary}
\author{Dirk Schuricht}
\affiliation{Institute for Theory of Statistical Physics, 
RWTH Aachen, 52056 Aachen, Germany}
\affiliation{JARA-Fundamentals of Future Information Technology}
\date{\today}
\pagestyle{plain}

\begin{abstract}
  We study the low-energy limit of a quarter-filled one-dimensional Mott
  insulator. We analytically determine the local density of states in the
  presence of a strong impurity potential, which is modeled by a boundary.  To
  this end we calculate the Green function using field theoretical methods.
  The Fourier transform of the local density of states shows signatures of a
  pinning of the spin-density wave at the impurity as well as several
  dispersing features at frequencies above the charge gap. These features can
  be interpreted as propagating spin and charge degrees of freedom. Their
  relative strength can be attributed to the ``quasi-fermionic'' behavior of
  charge excitations with equal momenta. Furthermore, we discuss the effect of
  bound states localized at the impurity. Finally, we give an overview of the 
  local density of states in various one-dimensional systems and discuss
  implications for scanning tunneling microscopy experiments.
\end{abstract}
\pacs{68.37.Ef, 71.10.Pm, 72.80.Sk}
\maketitle

\section{Introduction}
Over the past decades scanning tunneling microscopy (STM) and spectroscopy
techniques have been established as an important experimental tool to study
strongly correlated electron systems with a spatial resolution down to the
atomic scale.  In these experiments the tunneling current between the sample
and the STM tip is measured as a function of its position and the applied
voltage. The tunneling current is directly related~\cite{Fischer-07} to the
local density of states (LDOS) in the sample; thus STM experiments provide a
tool to investigate the spatial dependence of the LDOS, which may arise for
example due to impurities.  These spatial modulations can be analyzed in terms
of the Fourier transform of the tunneling conductance which is directly
proportional to the Fourier transform of the LDOS. Using this method of
analyzing STM data one can infer informations about the bulk state of matter
like the properties of its quasiparticle excitations.  For example, this has
been used to study high-temperature superconductors~\cite{Hoffman-02} as well
as carbon nanotubes~\cite{Lee-04}.

Theoretical studies of the LDOS and STM in the presence of boundaries have
been carried out in particular for one-dimensional systems including Luttinger
liquids~\cite{FabrizioGogolin95,Meden-00,Kivelson-03,Kakashvili-06,SchneiderEggert10},
open Hubbard chains~\cite{Beduerftig-98,Meden-00}, and charge-density wave
(CDW) states~\cite{SEJF08,SEJF11}.  In these systems the coupling to
impurities is relevant~\cite{KaneFisher92prl,Giamarchi04,SEJF11}, which leads,
at sufficiently low energies, to an effective cutting of the system into
disconnected pieces.  In this sense the effect of an impurity can be modeled 
by a boundary condition. Previous studies of the Fourier transform of the LDOS in
both gapless~\cite{Kivelson-03,Kakashvili-06,SchneiderEggert10} and
gapped~\cite{SEJF08,SEJF11} systems revealed a pinning of the CDW at the
impurity as well as individually propagating spin and charge degrees of
freedom.

Here we study a quarter-filled one-dimensional Mott insulator in the presence
of a strong impurity potential. It has been established~\cite{Giamarchi97,Giamarchi04} that
for sufficiently strong interactions double Umklapp processes are relevant and
generate a gap in the charge sector. As a microscopic realization one may
think of a quarter-filled Hubbard model extended by including nearest-neighbor
interactions, which possesses~\cite{MilaZotos93} an insulating phase for
sufficiently strong repulsions.  Experimentally these systems are relevant for
the description of organic quasi-one-dimensional conductors, i.e. the
Bechgaard and Fabre salts~\cite{Seo-04}.

In this article we study the low-energy behavior of the LDOS by employing the
field theoretical description of the Mott insulator. The spectrum consists of
massless spin excitations (spinons) carrying spin $\pm 1/2$ but no charge and
massive charge excitations (solitons and antisolitons) carrying charge $\pm
e/2$ but no spin. In this framework the impurity is modeled by a boundary for
the collective excitations.  In particular, we will focus on signatures of
dispersing quasiparticles in the spatial Fourier transform of the LDOS, as
they have been clearly observed in similar studies of Luttinger
liquids~\cite{Kivelson-03,Kakashvili-06,SchneiderEggert10} and half-filled
Mott insulators as well as CDW states~\cite{SEJF08,SEJF11}.  The spectral
function of the translationally invariant, quarter-filled Mott insulator has
been previously derived~\cite{EsslerTsvelik02prl} by Essler and Tsvelik. From
the quantum numbers stated above it is clear that an electron will
fractionalize into at least one spinon and two antisolitons.  Thus the
spectral function exhibits a featureless scattering continuum of at least
three excitations, which is also consistent with experimental
results~\cite{Zwick-97} from angle-resolved photoelectron spectroscopy on
Bechgaard salts.

This article is organized as follows. In Sec.~\ref{sec:model} we will briefly
discuss the field theoretical description of the quarter-filled Mott
insulator. In Sec.~\ref{sec:GF} we present our results for the single-particle
Green function. This will then be used to derive an analytic expression for
the LDOS, which we will discuss in detail in Sec.~\ref{sec:LDOS}. Our main
result is presented in Eq.~\eqref{eq:LDOSresult} as well as
Figs.~\ref{fig:plot1} and~\ref{fig:plot2}, which show the Fourier transform of
the LDOS in the presence of an impurity.  Surprisingly, we find two dispersing
features which follow the dispersion relations \eqref{eq:DRE1c} and
\eqref{eq:DRE1cs} respectively. These features can be interpreted as arising
from an antisoliton or a spinon-antisoliton pair propagating between the
position of the STM tip and the impurity.  In addition, we observe a
non-dispersing singularity at momentum $Q=2k_\mathrm{F}$ which is indicative
of a pinning of a spin-density wave (SDW) at the boundary. A detailed analysis
further reveals dispersing features originating in propagating spinons and
antisoliton pairs.  The suppression of these features is attributed to the
``quasi-fermionic'' behavior of antisolitons with equal momenta. Finally we
explain the different findings in the spectral function and the LDOS.  In
Sec.~\ref{sec:bbs} we discuss the effect of possible boundary bound states on
the LDOS and in Sec.~\ref{sec:HFMI} we compare our results to the half-filled Mott 
insulator studied~\cite{SEJF08,SEJF11} previously. Finally, in Sec.~\ref{sec:discussion} we set 
our results in the context of other one-dimensional systems and discuss 
implications for STM experiments on quasi-one-dimensional materials.
Readers mainly interested in the qualitative features of the LDOS 
may start with this section and study the detailed results afterwards. 
Technical details have been moved to the appendix.

\section{The model}
\label{sec:model}
As the underlying microscopic model for the quarter-filled Mott insulator one
may start with the extended Hubbard model~\cite{Giamarchi04}
\begin{equation}
  \begin{split}
  H_\mathrm{Hubbard}=&-t\sum_{j,\sigma}
  \left[c^\dagger_{j,\sigma}c_{j+1,\sigma}+
    c^\dagger_{j+1,\sigma}c_{j,\sigma}\right]\\
  &+U\sum_j n_{j,\uparrow}\,n_{j,\downarrow}+V\sum_j n_{j}\,n_{j+1},
  \label{eq:extHM}
  \end{split}
\end{equation}
where $c^\dagger_{j,\sigma}$ and $c_{j,\sigma}$ are electron creation and
annihilation operators at site $j$ with spin $\sigma=\uparrow,\downarrow$,
$n_{j,\sigma}=c^\dagger_{j,\sigma}c_{j,\sigma}$, and
$n_j=n_{j,\uparrow}+n_{j,\downarrow}$. At quarter filling and for sufficiently
large on-site and nearest-neighbor repulsions $U,V\ge 4t$ this model possesses
a Mott insulating phase~\cite{MilaZotos93} which can be thought of as a
microscopic realization of the field-theoretical system we will study below.
The extended Hubbard model \eqref{eq:extHM} can be used as an effective model
for the description of Bechgaard salts~\cite{Seo-04}; its spectral and
transport properties have been investigated in detail using various
techniques~\cite{Giamarchi04,Controzzi-01}.

Instead of working with the microscopic model \eqref{eq:extHM} we will study
here the corresponding low-energy field theory. The continuum description is
obtained by focusing on the degrees of freedom around the Fermi points $\pm
k_\mathrm{F}$. We introduce slowly varying right- and left-moving Fermi fields
as
\begin{equation}
\label{eq:lowenergy}
\frac{c_{j,\sigma}}{\sqrt{a_0}}\rightarrow\Psi_{\sigma}(x)=
e^{\ii k_\mathrm{F} x} R_\sigma(x)+
e^{-\ii k_\mathrm{F} x} L_\sigma(x),
\end{equation}
and similarly for the electron creation operators. Here $a_0$ denotes the
lattice spacing, $x=ja_0$, and the Fermi momentum is given by
$k_\mathrm{F}=\pi/4a_0$ at quarter filling.  The impurity potential is assumed
to be strong such that we can model it by a boundary condition on the
continuum electron field
\begin{equation}
  \Psi_\sigma(x=0)=0.
  \label{eq:BCPsi}
\end{equation}
Using standard bosonization of the right- and left-moving Fermi fields it was 
shown~\cite{Giamarchi97,Giamarchi04} that double Umklapp processes result in a
cosine term in the charge sector, which is relevant at low energies and generates
a gap. Thus the effective low-energy Hamiltonian is given by~\cite{endnote1}
\begin{eqnarray}
H&=&H_\mathrm{c}+H_\mathrm{s},\label{eq:hamiltonian}\\
H_\mathrm{c}&=&\frac{v_\mathrm{c}}{16\pi}\int_{-\infty}^0 dx 
\Bigl[\bigr(\partial_x\Phi_\mathrm{c}\bigr)^2+
\bigr(\partial_x\Theta_\mathrm{c}\bigr)^2\Bigr]\nonumber\\*
&&-\frac{g_\mathrm{c}}{(2\pi)^2}\int_{-\infty}^0 dx\,
\cos\bigl(\beta\Phi_\mathrm{c}),
\label{eq:chargehamiltonian}\\
H_\mathrm{s}&=&\frac{v_\mathrm{s}}{16\pi}\int_{-\infty}^0 dx 
\Bigl[\bigr(\partial_x\Phi_\mathrm{s}\bigr)^2+
\bigr(\partial_x\Theta_\mathrm{s}\bigr)^2\Bigr],
\label{eq:spinhamiltonian}
\end{eqnarray}
where, as a consequence of \eqref{eq:BCPsi}, the canonical Bose fields
$\Phi_\mathrm{c,s}$ are subject to the hard-wall boundary conditions
\begin{equation}
\Phi_\mathrm{c,s}(x=0)=0.
\label{eq:hardwall}
\end{equation} 
The fields $\Theta_\mathrm{c,s}$ are dual to $\Phi_\mathrm{c,s}$.  The charge
and spin velocities $v_\mathrm{c,s}$, the Luttinger parameter $\beta$, and the
coupling constant $g_\mathrm{c}$ are functions of the hopping and interactions
in an underlying microscopic model.  Within the field theory they can be
viewed as phenomenological parameters.  Experimental
estimates~\cite{Schwartz-98} for the Luttinger parameter in the Bechgaard
salts yield $\beta^2\approx 0.9$.

In the bosonization procedure leading to the effective low-energy Hamiltonian
\eqref{eq:hamiltonian} the explicit relation between the Fermi and the Bose
fields is given by~\cite{EsslerTsvelik02prl,EsslerKonik05}
\begin{eqnarray}
R^\dagger_\sigma&=&\frac{\eta_\sigma}{\sqrt{2\pi}}\,
\exp\biggl[\frac{\ii}{4}\biggl(\frac{\beta}{2}\Phi_\mathrm{c}
+\frac{2}{\beta}\Theta_\mathrm{c}\biggr)\biggr]\nonumber\\*
&&\qquad\qquad\times\exp\biggl[\frac{\ii}{4}f_\sigma\bigl(\Phi_\mathrm{s}
+\Theta_\mathrm{s}\bigr)\biggr],\label{eq:bosonizationR}\\*
L^\dagger_\sigma&=&\frac{\eta_\sigma}{\sqrt{2\pi}}\,
\exp\biggl[-\frac{\ii}{4}\biggl(\frac{\beta}{2}\Phi_\mathrm{c}
-\frac{2}{\beta}\Theta_\mathrm{c}\biggr)\biggr]\nonumber\\*
&&\qquad\qquad\times\exp\biggl[-\frac{\ii}{4}f_\sigma\bigl(\Phi_\mathrm{s}
-\Theta_\mathrm{s}\bigr)\biggr],\label{eq:bosonizationL}
\end{eqnarray}
where the Klein factors $\eta_\sigma$ satisfy the anticommutation relations
$\anticomm{\eta_\sigma}{\eta_\sigma'}=2\delta_{\sigma\sigma'}$ and
$f_\uparrow=1=-f_\downarrow$.

The charge excitations are described by the sine-Gordon model on the half-line
(\ref{eq:chargehamiltonian}). For $\beta<1$ the cosine term is relevant and
generates a gap. The excitations are massive solitons and antisolitons which
carry charges $\pm e/2$ respectively. They possess a relativistic dispersion
relation; we parametrize their energy and momentum in the usual way using the
rapidity $\theta$ by
\begin{equation}
  E=\Delta\cosh\theta,\quad P=\frac{\Delta}{v_\mathrm{c}}\sinh\theta.
  \label{eq:EP}
\end{equation}
The soliton mass $\Delta$ is a function~\cite{Zamolodchikov95} of the bare
parameters $g_\mathrm{c}$ and $\beta$ but will be viewed here as a
phenomenological parameter replacing the coupling constant $g_\mathrm{c}$. In
the regime $1/2\le\beta^2$ solitons and antisolitons are the only excitations
in the charge sector; for $\beta^2<1/2$ propagating breather
(soliton-antisoliton) bound states exist as well. At the Luther-Emery point
(LEP) $\beta^2=1/2$ the charge sector is equivalent to a free massive Dirac
theory~\cite{LutherEmery74}. The sine-Gordon model
(\ref{eq:chargehamiltonian}) with the boundary condition \eqref{eq:hardwall}
is integrable\cite{GhoshalZamolodchikov94,MacIntyre95}, which we will use
below to calculate the necessary correlation functions of right- and
left-moving Fermi fields. The spin sector (\ref{eq:spinhamiltonian}) describes
massless relativistic spinon excitations which propagate with velocity
$v_\mathrm{s}$ and carry spin $\pm 1/2$. We have already assumed spin
rotational invariance, i.e. a possible Luttinger parameter in the spin sector
is set to one.

We note that the half-filled Mott insulator studied previously~\cite{SEJF08,SEJF11} 
possesses the same effective low-energy Hamiltonian 
\eqref{eq:hamiltonian}--\eqref{eq:spinhamiltonian}. The difference to the quarter-filled case
studied in this article is given~\cite{EsslerKonik05} by the bosonization relations 
\eqref{eq:bosonizationR} and \eqref{eq:bosonizationL} for the right- and left-moving Fermi fields.
This leads to differing Green functions as well as LDOS. We will proceed with the 
presentation of the results for the quarter-filled Mott insulator and present a detailed
comparison to the half-filled system in Sec.~\ref{sec:HFMI}.

\section{Green function}
\label{sec:GF}
In order to derive the LDOS below we calculate the time-ordered
Green function in Euclidean space,
\begin{equation}
\label{eq:GF}
G_{\sigma\sigma'}(\tau,x_1,x_2)=
-\bra{0_\mathrm{b}}\mathcal{T}_\tau\,\Psi_\sigma(\tau,x_1)\,
\Psi_{\sigma'}^\dagger(0,x_2)\ket{0_\mathrm{b}},
\end{equation}
where $\ket{0_\mathrm{b}}$ is the ground state of \eqref{eq:hamiltonian} in
the presence of the boundary and $\tau=\ii t$ denotes imaginary time. At low
energies we can linearize around the Fermi points $\pm k_\mathrm{F}$, which
results in
\begin{equation}
\begin{split}
\label{eq:GFlowenergy}
G_{\sigma\sigma'}=&
e^{\ii k_\mathrm{F}(x_1-x_2)}\,G^{RR}_{\sigma\sigma'}
+e^{-\ii k_\mathrm{F}(x_1-x_2)}\,G^{LL}_{\sigma\sigma'}\\
&+e^{\ii k_\mathrm{F}(x_1+x_2)}\,G^{RL}_{\sigma\sigma'}
+e^{-\ii k_\mathrm{F}(x_1+x_2)}\,G^{LR}_{\sigma\sigma'},
\end{split}
\end{equation}
where e.g. $G^{RL}_{\sigma\sigma'}=-\bra{0_\mathrm{b}}\mathcal{T}_\tau\,
R_\sigma(\tau,x_1)\,L^\dagger_{\sigma'}(0,x_2)\ket{0_\mathrm{b}}$. For the
calculation of the LDOS we have to set $x_1=x_2$ below. In particular, we will
focus on the spatial Fourier transform of the LDOS as physical properties can
be more easily identified. The four terms in the decomposition
\eqref{eq:GFlowenergy} then contribute in different regions in momentum space,
i.e. $G^{RL}_{\sigma\sigma'}$ and $G^{LR}_{\sigma\sigma'}$ contribute for
$Q\approx \pm 2k_\mathrm{F}$ while $G^{RR}_{\sigma\sigma'}$ and
$G^{LL}_{\sigma\sigma'}$ contribute for $Q\approx 0$. In the translationally
invariant system right- and left-moving fields in the spin sector decouple
which results in $G^{RL}_{\sigma\sigma'}=G^{LR}_{\sigma\sigma'}=0$. The
presence of a boundary couples right and left sectors and concomitantly the
Fourier transform of the Green function (\ref{eq:GFlowenergy}) acquires a
non-zero component at $Q\approx \pm 2k_\mathrm{F}$. We will focus on the
$2k_\mathrm{F}$-part of the Green function and the LDOS in the following,
since this provides a particularly clean way of investigating boundary
effects. We note that the other components of the Green function
\eqref{eq:GFlowenergy} can be calculated in the same way.

Due to the spin-charge separated Hamiltonian \eqref{eq:hamiltonian} the Green
function $G_{\sigma\sigma'}^{RL}$ factorizes into a product of correlation
functions in the spin and charge sectors. The correlation functions in the
spin sector can be straightforwardly derived using standard methods like
boundary conformal field theory~\cite{DiFrancescoMathieuSenechal97}.  On the
other hand, the integrability of the sine-Gordon model on the half-line
\eqref{eq:spinhamiltonian} enables us to calculate correlation functions in
the charge sector using the boundary state
formalism~\cite{GhoshalZamolodchikov94} together with a form-factor
expansion~\cite{Smirnov92book,Lukyanov95,EsslerKonik05}.  This yields the
following result for the $2k_\mathrm{F}$-component of the Green function
\begin{equation}
  \begin{split}
  &G^{RL}_{\sigma\sigma'}(\tau,x_1,x_2)\\
  &\qquad=-\frac{1}{2\pi}
  \frac{\delta_{\sigma\sigma'}}{\sqrt{v_\mathrm{s}\tau-\ii(x_1+x_2)}}
  \sum_{k=0}^\infty \mathcal{G}_k(\tau,x_1,x_2).
  \end{split}
  \label{eq:GRL}
\end{equation}
The infinite series originates from the applied form-factor expansion, which
constitutes an expansion of correlation functions in the number of
contributing solitons and antisolitons.  In App.~\ref{app:GF} we derive
explicit expressions for the first three terms $\mathcal{G}_k$ which
correspond to two-particle processes in the charge sector. We note that 
\eqref{eq:GRL} is valid at zero temperature. The effect of small temperatures
$T\ll\Delta$ on the gapless spin sector can be incorporated using 
conformal field theory~\cite{DiFrancescoMathieuSenechal97} as discussed 
for the half-filled Mott insulator in Ref.~\onlinecite{SEJF11}. In the next section
we use the result \eqref{eq:GRL} to derive the Fourier transform of the LDOS.

\section{Local density of states}
\label{sec:LDOS}
The LDOS can now be calculated directly from the Green function.  In order to
analyze the physical properties it is useful to consider the spatial Fourier
transform of the LDOS as features like dispersing quasiparticles can be more
easily identified.  This technique was previously applied to the LDOS of
Luttinger liquids~\cite{Kivelson-03,Kakashvili-06,SchneiderEggert10} as well
as CDW states and half-filled Mott insulators~\cite{SEJF08,SEJF11}. The
Fourier transform of the LDOS for positive energies $E>0$ is given
by~\cite{endnote2}
\begin{equation}
\begin{split}
N(E,Q)=-\frac{1}{2\pi}
\int_{-\infty}^0&dx\int^\infty_{-\infty} dt\,e^{\ii(E t-Qx)}\\
&\times
G_{\sigma\sigma}(\tau>0,x,x)\Big|_{\tau\rightarrow\ii t+\delta}.
\label{eq:defNp}
\end{split}
\end{equation}
Here the Green function has been analytically continued to real times and we
have taken the limit $x_1\rightarrow x_2\equiv x$. We note that due to the
assumed spin rotational invariance the LDOS does not depend on $\sigma$. As
mentioned before, we will concentrate on the $2k_\mathrm{F}$-component as it
vanishes in the absence of the boundary and hence offers a particularly clean
way of investigating boundary effects.  For $Q\approx 2k_\mathrm{F}$ only
$G^{RL}$ contributes and starting from \eqref{eq:GRL} we arrive at
our main result ($|q|\ll 2k_\mathrm{F}$)
\begin{eqnarray}
  N(E,2k_\mathrm{F}+q)\!\!&=&\!\!
  -\Theta(E-2\Delta)\sum_{k=0}^2N_k(E,2k_\mathrm{F}+q),
  \qquad\label{eq:LDOSresult}\\
  N_k(E,2k_\mathrm{F}+q)\!\!&=&\!\!
  Z_2\frac{e^{-\ii\pi/4}\sqrt{v_\mathrm{s}}}{2\pi^{3/2}}\nonumber\\*
  & &\hspace{-15mm}\times\int\frac{d\theta_1d\theta_2}{(2\pi)^2}
  \frac{\Theta(E-\Delta\sum_i\cosh\theta_i)}
  {\sqrt{E-\Delta\sum_i\cosh\theta_i}}\nonumber\\*
  & &\hspace{-5mm}\times\frac{h_k(\theta_1,\theta_2)}
  {v_\mathrm{s}q_k-2(E-\Delta\sum_i\cosh\theta_i)+\ii\delta},
  \label{eq:LDOSresult2}
\end{eqnarray}
where
\begin{eqnarray*}
  h_0(\theta_1,\theta_2)\!\!&=&\!\!\frac{1}{2}\,|G(\theta_1-\theta_2)|^2,\\
  h_1(\theta_1,\theta_2)\!\!&=&\!\!
  K(\theta_1+\ii\tfrac{\pi}{2})\,e^{\theta_1/4}\,
  G(\theta_2-\theta_1)\,G(\theta_1+\theta_2)^*,\\
  h_2(\theta_1,\theta_2)\!\!&=&\!\!
  -\frac{1}{2}\,e^{(\theta_1+\theta_2)/4}\,\prod_{i=1}^2K(\theta_i+\ii\tfrac{\pi}{2})\\*
  & &\hspace{-15mm}\times \frac{\sinh\frac{\theta_1+\theta_2+\ii\pi}{\xi}}
  {\sinh\frac{\theta_1+\theta_2}{\xi}}\,S_0(\theta_1+\theta_2+\ii\pi)\,|G(\theta_1-\theta_2)|^2,
\end{eqnarray*}
as well as $q_0=q$, $q_1=q-\frac{2\Delta}{v_\mathrm{c}}\sinh\theta_1$, and
$q_2=q-\frac{2\Delta}{v_\mathrm{c}}\sum_i\sinh\theta_i$. Explicit integral
representations for the scattering matrix $S_0(\theta)$, the boundary
reflection matrix $K(\theta)$, and the function $G(\theta)$ are given in
App.~\ref{app:ff}. Physically $S_0(\theta)$ describes the scattering of two
antisolitons with relative momentum corresponding to the rapidity $\theta$, 
$K(\theta)$ is the reflection amplitude of an antisoliton with rapidity $\theta$ 
off the boundary [and thus incorporates the information on the boundary 
condition \eqref{eq:hardwall}], and $G(\theta)$ appears in the matrix element
[i.e. form factor] of the charge part of \eqref{eq:bosonizationR} and \eqref{eq:bosonizationL}
with two-antisoliton states [see \eqref{eq:ff}]. The parameter $\xi$ is defined as
$\xi=\beta^2/(1-\beta^2)$. The normalization
constant\cite{LukyanovZamolodchikov01} $Z_2$ has dimension
$Z_2\propto\Delta^{\beta^2/16+1/\beta^2}$; thus in order to obtain
dimensionless quantities we multiply the LDOS by
$\Delta^{3/2-\beta^2/16-1/\beta^2}/\sqrt{v_\mathrm{s}}$ in all figures. The
singularity of the integrands in \eqref{eq:LDOSresult2} is smeared out by
taking $\delta$ small but finite; we choose $\delta=0.01$ unless stated
otherwise. This mimics the broadening of singularities in experiments due to
the instrumental resolution and the finite temperature. The LDOS
\eqref{eq:LDOSresult} vanishes for $E<2\Delta$ as the fractionalization of an
electron creates at least two antisolitons.

The three terms \eqref{eq:LDOSresult2} constitute the leading contribution
from the form-factor expansion. As can be seen from the double integrals they
all originate in two-particle processes in the charge sector. The number of
boundary reflection matrices $K(\theta)$ indicates that $N_{k}$ involves $k$
interactions with the boundary, i.e. the terms \eqref{eq:LDOSresult2} contain
no, one, or two reflections of charge excitations.  The sub-leading
contributions involve at least three particles in the charge sector which
either come from a higher number of particles in the intermediate state or
from higher-order processes due to the boundary.  Similar terms were
thoroughly analyzed for the LDOS in a CDW state at the LEP~\cite{SEJF11} as
well as the spectral function in the Ising model with a boundary magnetic
field~\cite{SE07}; they were found to be negligible in both cases.  We further
note that the suppression of sub-leading terms in the form-factor expansion
for bulk two-point functions is a well-known feature of massive integrable
field theories~\cite{CardyMussardo93,EsslerKonik05}.  We thus expect the
higher-order corrections to \eqref{eq:LDOSresult} to be negligible in the
low-energy regime we consider here.

\begin{figure}[tb]
\centering
\includegraphics[scale=0.34,clip=true]{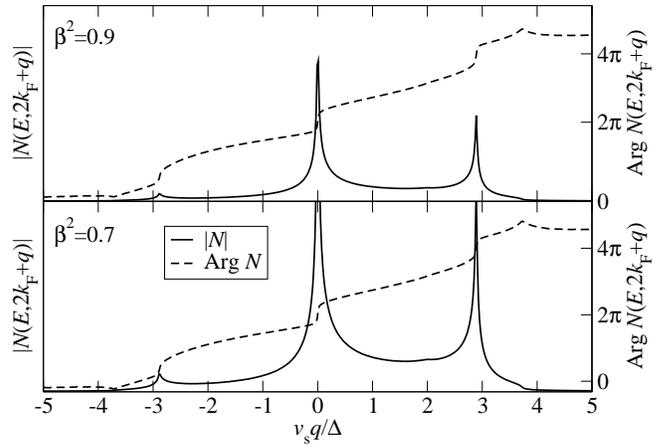}
\caption{Constant energy scan for $E=3\Delta$:
  $|N(E,2k_\mathrm{F}+q)|$ and $\text{Arg}\,N(E,2k_\mathrm{F}+q)$ for
  $v_\mathrm{c}=1.2\,v_\mathrm{s}$ and $\beta^2=0.9$ (upper panel) and
  $\beta^2=0.7$ (lower panel). Both plots are shown on the same scale. We
  observe a peak at $q=0$ and dispersing features at $q=\pm
  2\sqrt{(E-\Delta)^2-\Delta^2}/v_\mathrm{c}$ as well as $q=\pm
  2\sqrt{E^2-4\Delta^2}/v_\mathrm{c}$. For $q<0$ the dispersing features
  are strongly suppressed.}
\label{fig:plot1}  
\end{figure}
\begin{figure}[tb]
\centering
\includegraphics[scale=0.34,clip=true]{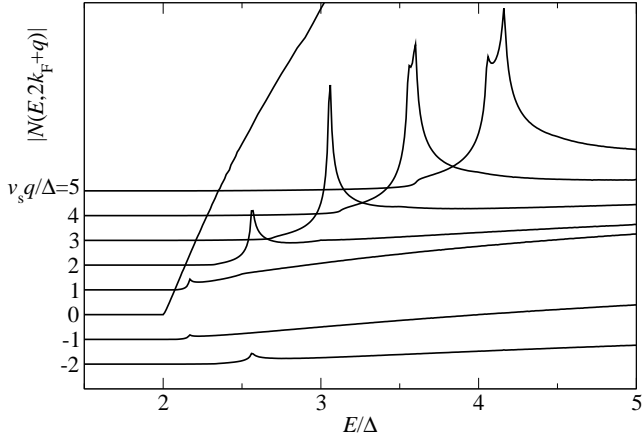}
\caption{$|N(E,2k_\mathrm{F}+q)|$ for $\beta^2=0.9$ and
  $v_\mathrm{c}=1.2\,v_\mathrm{s}$. The curves are constant $q$-scans which
  have been offset along the y-axis by a constant with respect to one another.
  The LDOS is dominated by the peak at $q=0$. We further observe dispersing
  features at $E_\mathrm{1c}=\Delta+\sqrt{\Delta^2+(v_\mathrm{c}q/2)^2}$ as
  well as $E_\mathrm{1cs}=\Delta+\Delta\sqrt{1-(v_\mathrm{s}/v_\mathrm{c})^2}
  +v_\mathrm{s}q/2$ for $q>q_0$.}
\label{fig:plot2}
\end{figure}

The constant energy scan of the Fourier transform of the LDOS
\eqref{eq:LDOSresult} is shown in Fig.~\ref{fig:plot1}. We observe a
singularity at $q=0$, i.e.  $Q=2k_\mathrm{F}$, which is indicative of the
pinning of the $2k_\mathrm{F}$-SDW at the boundary (see Sec.~\ref{sec:LDOS0}).
Furthermore there is a peak at $q>0$. The constant momentum scans shown in
Fig.~\ref{fig:plot2} reveal that this peak shows dispersing behavior and
splits above a critical momentum $q>q_0$ [$q_0$ will be defined in
Eq.~\eqref{eq:defq0}]. More precisely these features follow the dispersion
relations \eqref{eq:DRE1c} and \eqref{eq:DRE1cs} respectively (see
Sec.~\ref{sec:LDOS1}).  The existence of dispersing features is at first
surprising since the spectral function~\cite{EsslerTsvelik02prl} of the
translationally invariant system does not show any dispersing quasiparticles.
The processes contributing to the LDOS \eqref{eq:LDOSresult} can be thought of
as arising from creating an electron at a position $x$, the fractionalization
of this electron into at least one spinon and two antisolitons, the
propagation of these elementary excitations through the system, and their
subsequent recombination at the position $x$.  This allows a natural
interpretation of the leading terms in the form-factor expansion
\eqref{eq:LDOSresult2} in terms of propagating antisolitons and spinons.  In
the next sub-sections we will analyze the different terms
\eqref{eq:LDOSresult2} separately. Afterwards we comment on the different
findings in the spectral function and the LDOS.

\subsection{No reflections in the charge sector: 
$\boldsymbol{N_0(E,2k_\mathrm{F}+q)}$}
\label{sec:LDOS0}
\begin{figure}[tb]
  \centering \includegraphics[scale=0.34,clip=true]{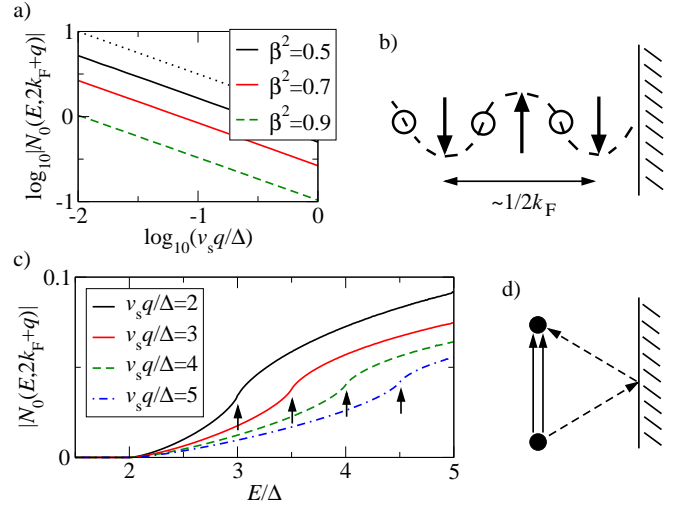}
\caption{(Color online) $N_0(E,2k_\mathrm{F}+q)$ for 
  $v_\mathrm{c}=1.6\,v_\mathrm{s}$. a) Logarithmic plot close to the
  singularity at $q=0$ for $\delta=0.001$. The dotted line is a guide to the
  eye; it has gradient $-1/2$. b) Sketch~\cite{Giamarchi04,SeoFukuyama97} of
  the $2k_\mathrm{F}$-SDW pinned at the boundary. c)
  $|N_0(E,2k_\mathrm{F}+q)|$ for $\beta^2=0.9$ and 
  different momenta. We observe a dispersing
  feature at $E_\mathrm{s}=2\Delta+v_\mathrm{s}q/2$ indicated by small arrows.
  d) Sketch of the process giving rise to the propagating feature. The
  electron decomposes into two zero-momentum antisolitons (solid lines) and one
  spinon (dashed line) reflected at the boundary.}
\label{fig:plot3}
\end{figure}

Let us start with the analysis of the first term in the form-factor expansion
\eqref{eq:LDOSresult}, which is shown in Fig.~\ref{fig:plot3}. We first note
that this term is dominated by a singularity at $q=0$, i.e. $Q=2k_\mathrm{F}$.
This singularity is due to the pinning of the $2k_\mathrm{F}$-SDW at the
boundary.  As $k_\mathrm{F}=\pi/4a_0$ we find a spatial modulation with
periodicity $2\pi/2k_\mathrm{F}=4a_0$ as
sketched~\cite{Giamarchi04,SeoFukuyama97} in Fig.~\ref{fig:plot3}.b.  Close to
$q=0$ the first term in the form-factor expansion and thus the whole LDOS
behaves as~\cite{endnote3}
\begin{equation}
  N(E,2k_\mathrm{F}+q)\sim N_0(E,2k_\mathrm{F}+q)
  \sim\frac{1}{\sqrt{v_\mathrm{s}q}}.
\end{equation}
We note that the exponent is independent of the Luttinger parameter $\beta$
(see Fig.~\ref{fig:plot3}.a). This singularity is similar to the ones observed
in the LDOS of Luttinger liquids~\cite{Kivelson-03,SchneiderEggert10} as well
as CDW states~\cite{SEJF08,SEJF11} where, however, the exponents depend on the
interaction strength and thus the Luttinger parameter.

Furthermore, $N_0$ possesses a weak dispersing feature following
(see~\cite{endnote4} Fig.~\ref{fig:plot3}.c)
\begin{equation}
  E_\mathrm{s}(q)=2\Delta+\frac{v_\mathrm{s}q}{2},\quad q>0.
  \label{eq:DREs}
\end{equation}
As already discussed, the created electron fractionalizes into two
antisolitons and (at least) one spinon which can propagate through the system.
In this way of thinking the dispersing feature at $E_\mathrm{s}$ arises from
two antisolitons with zero momentum (solid lines in Fig.~\ref{fig:plot3}.d)
contributing an energy $2\Delta$ and a massless soliton with momentum $q$
(dashed line in Fig.~\ref{fig:plot3}.d).  The appearance of $v_\mathrm{s}/2$
in $E_\mathrm{s}$ is due to the fact that the spinon has to propagate to the
boundary and back, thus covering the distance $2x$ in time $t$.

\subsection{One reflection in the charge sector: 
  $\boldsymbol{N_1(E,2k_\mathrm{F}+q)}$}
\label{sec:LDOS1}
\begin{figure}[tb]
\centering
\includegraphics[scale=0.34,clip=true]{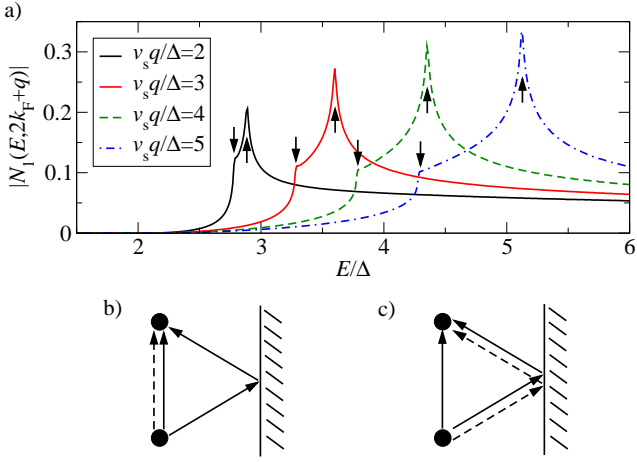}
\caption{(Color online) a) $|N_1(E,2k_\mathrm{F}+q)|$ for $\beta^2=0.9$, 
  $v_\mathrm{c}=1.6\,v_\mathrm{s}$, and different momenta. We observe dispersing 
  features at $E_\mathrm{1c}=\Delta+\sqrt{\Delta^2+(v_\mathrm{c}q/2)^2}$ indicated by
  up-arrows and
  $E_\mathrm{1cs}=\Delta+\Delta\sqrt{1-(v_\mathrm{s}/v_\mathrm{c})^2}
  +v_\mathrm{s}q/2$ indicated by down-arrows. b) Process resulting in
  $E_\mathrm{1c}$. The electron decomposes into one antisoliton and spinon
  with zero momentum as well as one antisoliton reflected at the boundary. c)
  Process resulting in $E_\mathrm{1cs}$.}
\label{fig:plot4}
\end{figure}
The next term in the form-factor expansion \eqref{eq:LDOSresult} contains one
boundary reflection matrix $K(\theta)$. Thus it is natural to interpret it as
arising from processes involving the reflection of one antisoliton at the
boundary. In Fig.~\ref{fig:plot4}.a we show $N_1$ for different momenta. We
clearly observe a dispersing feature (indicated by up-arrows) following
\begin{equation}
  E_\mathrm{1c}(q)=\Delta+
  \sqrt{\Delta^2+\left(\frac{v_\mathrm{c}q}{2}\right)^2}.
  \label{eq:DRE1c}
\end{equation}
This feature originates in the process where one antisoliton propagates
with momentum $q$ while the other antisoliton and the spinon possess zero
momentum (see Fig.~\ref{fig:plot4}.b). The first term in $E_\mathrm{1c}$ is
the rest mass of the zero-momentum antisoliton while the second term is the
energy of the propagating antisoliton.

Furthermore, when $q$ exceeds the critical value $q_0$ a second dispersing
feature appears (indicated by down-arrows in Fig.~\ref{fig:plot4}.a) at
\begin{equation}
E_\mathrm{1cs}(q)=\Delta
+\Delta\sqrt{1-\left(\frac{v_\mathrm{s}}{v_\mathrm{c}}\right)^2}+
\frac{v_\mathrm{s}q}{2}.
\label{eq:DRE1cs}
\end{equation}  
The critical momentum $q_0$ can be determined from the condition that the spin
and charge excitations have the same group velocity
\begin{equation}
\frac{\partial E_\mathrm{1c}}{\partial q}\bigg|_{q=q_0}\stackrel{!}{=}
\frac{\partial E_\mathrm{s}}{\partial q}=\frac{v_\mathrm{s}}{2},
\end{equation}
which leads to
\begin{equation}
  q_0=\frac{2\Delta v_\mathrm{s}}{v_\mathrm{c}
    \sqrt{v_\mathrm{c}^2-v_\mathrm{s}^2}}.
  \label{eq:defq0}
\end{equation}
Thus this feature can be thought of as arising from an antisoliton with
momentum $q_0$ and a spinon carrying momentum $q-q_0$, while the second
antisoliton stays at rest (see Fig.~\ref{fig:plot4}.c).  The fact that the
propagating antisoliton and the spinon have the same group velocity results in
an increased probability that both simultaneously arrive at position $x$ after
reflection at the boundary.

The observed splitting of the quasiparticle peak is reminiscent of what is
found for the single-particle spectral function in the translationally
invariant, half-filled Mott insulator~\cite{Voit98} as well as the LDOS of a
CDW state in the presence of a boundary in the regime of attractive
interactions~\cite{SEJF08,SEJF11}.  The peak splitting is a consequence of the
curvature of the antisoliton dispersion (and thus of the charge gap) together
with the condition $v_\mathrm{c}>v_\mathrm{s}$.  Hence it cannot be observed
in the Luttinger liquid case~\cite{Kivelson-03,Kakashvili-06,SchneiderEggert10} where 
both sectors are massless.

\subsection{Two reflections in the charge sector: 
$\boldsymbol{N_2(E,2k_\mathrm{F}+q)}$}
\label{sec:LDOS2}
\begin{figure}[tb]
\centering
\includegraphics[scale=0.34,clip=true]{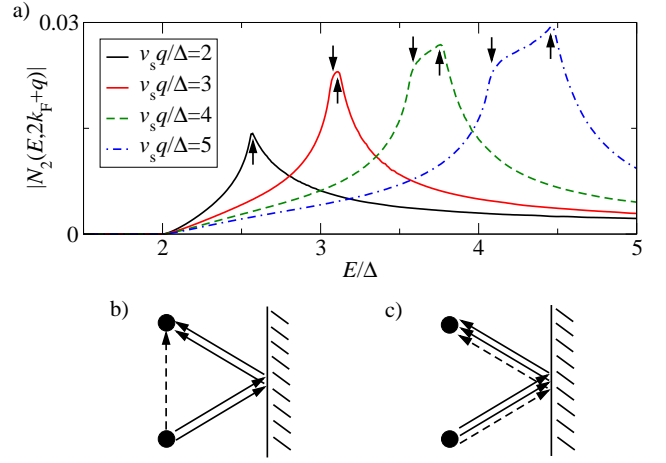}
\caption{(Color online) a) $|N_2(E,2k_\mathrm{F}+q)|$ for $\beta^2=0.9$,
  $v_\mathrm{c}=1.6\,v_\mathrm{s}$, and 
  different momenta. We observe dispersing features at
  $E_\mathrm{2c}=2\sqrt{\Delta^2+(v_\mathrm{c}q/4)^2}$ (up-arrows) and
  $E_\mathrm{2cs}=2\Delta\sqrt{1-(v_\mathrm{s}/v_\mathrm{c})^2}
  +v_\mathrm{s}q/2$ (down-arrows). Processes resulting in (b) $E_\mathrm{2c}$
  and in (c) $E_\mathrm{2cs}$.}
\label{fig:plot5}
\end{figure}
The leading term in second order in the boundary reflection matrix is $N_2$.
As in the previous term we find two dispersing features. First, indicated by
up-arrows in Fig.~\ref{fig:plot5}.a, there is a feature following
\begin{equation}
  E_\mathrm{2c}(q)=2\sqrt{\Delta^2+\left(\frac{v_\mathrm{c}q}{4}\right)^2}.
  \label{eq:DRE2c}
\end{equation}
This feature can be thought of as arising from the propagation of both
antisolitons with momenta $q/2$ and thus the same velocity, while the spinon
has zero momentum (see Fig.~\ref{fig:plot5}.b). The second feature is found at
\begin{equation}
E_\mathrm{2cs}(q)=
2\Delta\sqrt{1-\left(\frac{v_\mathrm{s}}{v_\mathrm{c}}\right)^2}+
\frac{v_\mathrm{s}q}{2},
\label{eq:DRE2cs}
\end{equation}  
provided $q$ exceeds the critical momentum $2q_0$. This feature is indicated
by down-arrows in Fig.~\ref{fig:plot5}.a. The interpretation is that both
antisolitons carry momentum $q_0$ while the excess momentum $q-2q_0$ is
carried by the spinon (see Fig.~\ref{fig:plot5}.c). 

We note that at finite momentum $q>0$ the term $N_1$ dominates the LDOS as can
be seen by comparing the scales in Figs.~\ref{fig:plot3}.c,
~\ref{fig:plot4}.a, and~\ref{fig:plot5}.a, respectively.  More importantly,
the dispersing features originating in $N_1$ are much more pronounced than
those resulting from $N_0$ and $N_2$.  We attribute this to the fact that, at
least in the naive interpretation of Figs.~\ref{fig:plot4}.b and c, the
antisolitons involved in processes contributing to $N_1$ possess different
momenta as one of them is at rest while the other propagates through the
system. In contrast, the dispersing features in $N_0$ and $N_2$ originate from
processes in which the two antisolitons possess equal momenta (see
Figs.~\ref{fig:plot3}.d as well as~\ref{fig:plot5}.b and c, respectively).
Thus their relative rapidity vanishes and the scattering matrix is
$S_0(\theta=0)=-1$ (see App.~\ref{app:ff}). The resulting ``quasi-fermionic''
behavior leads to a strong suppression of the dispersing features in $N_0$
and $N_2$ via the Pauli principle.

Finally let us comment on the different findings in the spectral
function~\cite{EsslerTsvelik02prl} of the translationally invariant system and
the LDOS \eqref{eq:LDOSresult} in the presence of a boundary.  The spectral
function is obtained from the Green function $G(\tau,x,0)$ via Fourier
transformation with respect to $t=-\ii\tau$ and $x$. The contributing processes
require the propagation of all excitations, and in particular both
antisolitons, from $(0,0)$ to $(t,x)$.  Thus one may expect a dispersing
feature at $E=2\sqrt{\Delta^2+(v_\mathrm{c}q/2)^2}$ from the antisolitons
propagating with the same momentum $q/2$. In addition, above the critical
momentum $2q_0=2\Delta
v_\mathrm{s}/v_\mathrm{c}\sqrt{v_\mathrm{c}^2-v_\mathrm{s}^2}$ a linearly
dispersing feature following
$E=2\Delta\sqrt{1-(v_\mathrm{s}/v_\mathrm{c})^2}+v_\mathrm{s}q$ is expected.
Indeed these are exactly the conditions~\cite{EsslerTsvelik02prl} for the
threshold below which the spectral function vanishes.  Above the threshold the
spectral function is rather featureless, in particular there are no dispersing peaks
associated with the propagation of charge or spin degrees of freedom. 
The absence of dispersing features may be attributed to the ``quasi-fermionic'' 
behavior of antisolitons, as all processes contributing to the spectral function
will contain antisolitons with equal momenta. In contrast, the processes contributing to the LDOS
\eqref{eq:defNp} are given by the fractionalization of an electron at position
$x$, the propagation of the two antisolitons and the spinon through the
system, and the subsequent recombination at the original position $x$. In
particular, processes where the momentum $q$ is solely carried by one of the
antisolitons or a spinon-antisoliton pair will contribute, resulting in the
dispersing features at $E_\mathrm{1c}$ and $E_\mathrm{1cs}$, respectively.
The observation of these features thus underlines that the spectral function
and the LDOS probe rather different physical processes. While the spectral 
function requires the propagation of all excitations between two points in the system, the 
LDOS also probes processes where only one or two excitations propagate
to the impurity and back.

\section{Local density of states in the presence of 
boundary bound states}
\label{sec:bbs}
Up to now we have concentrated on the boundary conditions \eqref{eq:BCPsi} for
the electron field or equivalently \eqref{eq:hardwall} for the canonical Bose
fields.  In this section we will consider more general phase shifts for the
electron fields which lead to arbitrary Dirichlet boundary conditions for the
Bose fields. As is well known, such boundary conditions can result in the
existence of boundary bound states (BBS's), which are
expected~\cite{SEJF08,SEJF11} to manifest themselves in the LDOS as features
within the spectral gap. Here we will focus on general Dirichlet boundary
conditions in the charge sector
\begin{equation}
  \Phi_\mathrm{c}(x=0)=\Phi_\mathrm{c}^0,\quad
  -\frac{\pi}{\beta}\le\Phi_\mathrm{c}^0\le\frac{\pi}{\beta},
  \label{eq:genDBC}
\end{equation}
while in the spin sector we still assume $\Phi_\mathrm{s}(x=0)=0$. In the
boundary state formalism~\cite{GhoshalZamolodchikov94} applied here the
information about the boundary condition \eqref{eq:genDBC} is encoded in the
boundary reflection matrix $K(\theta)$. In particular, the existence of a BBS
will show up as a pole in the physical strip
$0\le\mathrm{Im}\,\theta\le\pi/2$. As was shown in
Ref.~\onlinecite{MattssonDorey00} this will be the case provided
\begin{equation}
  -\frac{\pi}{\beta}<\Phi_\mathrm{c}^0<-\beta\pi;
\end{equation}
the pole at $\theta=\ii\gamma$ corresponds to a BBS with energy
\begin{equation}
  E_\mathrm{bbs}=\Delta\sin\gamma,\quad
  \gamma=\frac{\pi+\beta\Phi_\mathrm{c}^0}{2-2\beta^2}.
\end{equation}
When calculating the Green function the pole of the boundary reflection matrix
will lead to two additional terms, $\mathcal{G}_3$ and $\mathcal{G}_4$, in the
form-factor expansion \eqref{eq:GRL}. This in turn yields two additional terms
in the LDOS \eqref{eq:LDOSresult}.

\begin{figure}[tb]
\centering
\includegraphics[scale=0.34,clip=true]{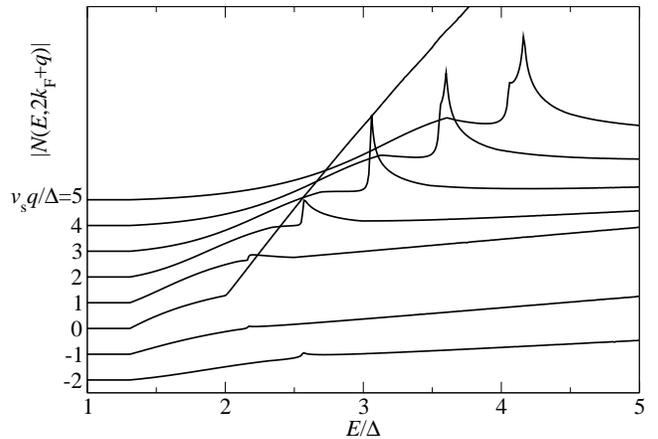}
\caption{$|N(E,2k_\mathrm{F}+q)|$ for $\beta^2=0.5$,
  $\Phi_\mathrm{c}^0=-4$, and $v_\mathrm{c}=1.2\,v_\mathrm{s}$. The curves are
  constant $q$-scans which have been offset along the y-axis by a constant
  with respect to one another. The existence of a BBS results in spectral
  weight within the two-soliton gap $E<2\Delta$.}
\label{fig:plot6}
\end{figure}
To be explicit, let us consider the LEP $\beta^2=1/2$ in the following. The
three leading terms in the form-factor expansion of the LDOS are still given
by \eqref{eq:LDOSresult2} where the boundary reflection matrix now depends
explicitly on $\Phi_\mathrm{s}^0$
\begin{equation}
  K(\theta)=
  \frac{\sin\bigl(\ii\frac{\theta}{2}-\frac{\Phi_\mathrm{c}^0}{\sqrt{8}}\bigr)}
  {\cos\bigl(\ii\frac{\theta}{2}+\frac{\Phi_\mathrm{c}^0}{\sqrt{8}}\bigr)}.
  \label{eq:KLEP}
\end{equation}
This obviously possesses a pole in the physical strip
$0\le\mathrm{Im}\,\theta\le\pi/2$ provided $\Phi_\mathrm{c}^0<-\pi/\sqrt{2}$
corresponding to the existence of a BBS.  The resulting two additional terms
in the LDOS read ($k=3,4$)
\begin{eqnarray}
  N_k(E,2k_\mathrm{F}+q)\!\!&=&\!\!
  Z_2\,\sqrt{v_\mathrm{s}}\,\cos\frac{\Phi_\mathrm{c}^0}{\sqrt{2}}\,
  \frac{e^{-\ii(\pi-\sqrt{2}\Phi_\mathrm{c}^0)/8}}
  {\pi^{3/2}}\nonumber\\*
  & &\hspace{-25mm}\times\int\frac{d\theta}{2\pi}
  \frac{\Theta(E-\Delta\cosh\theta-E_\mathrm{bbs})}
  {\sqrt{E-\Delta\cosh\theta-E_\mathrm{bbs}}}\nonumber\\*
  & &\hspace{-20mm}\times\frac{h_k(\theta)}
  {v_\mathrm{s}q_k-2(E-\Delta\cosh\theta-E_\mathrm{bbs})+
    \ii v_\mathrm{s}\kappa_\mathrm{bbs}},\qquad
  \label{eq:LDOSresult3}
\end{eqnarray}
where 
\begin{eqnarray*}
  h_3(\theta)\!\!&=&\!\!\biggl|\sinh\frac{\theta
    +\frac{\ii}{2}(\pi+\sqrt{2}\Phi_\mathrm{c}^0)}{2}\biggr|^2,\\
  h_4(\theta)\!\!&=&\!\!-K(\theta+\ii\tfrac{\pi}{2})\,e^{\theta/4}\,
  \sinh^2\frac{\theta-\frac{\ii}{2}(\pi+\sqrt{2}\Phi_\mathrm{c}^0)}{2},\\
  E_\mathrm{bbs}\!\!&=&\!\!-\Delta\sin\frac{\Phi_\mathrm{c}^0}{\sqrt{2}},
  \qquad\kappa_\mathrm{bbs}=-\frac{2\Delta}{v_\mathrm{c}}
  \cos\frac{\Phi_\mathrm{c}^0}{\sqrt{2}},
\end{eqnarray*}
as well as $q_3=q$ and $q_4=q-\frac{2\Delta}{v_\mathrm{c}}\sinh\theta$.  We
note that $\kappa_\mathrm{bbs}>0$ has the dimension of an inverse length. The
spatial dependence of the terms $\mathcal{G}_3$ and $\mathcal{G}_4$ (see
App.~\ref{app:GF}) is dominated by $\sim e^{\kappa_\mathrm{bbs}x}$ ($x<0$),
thus it is natural to interpret $1/\kappa_\mathrm{bbs}$ as the width of the
BBS.

The terms \eqref{eq:LDOSresult3} possess a threshold at
$E=\Delta+E_\mathrm{bbs}$ corresponding to the creation of the boundary bound
state and an additional antisoliton. Interestingly, although $N_3$ and $N_4$
separately possess jumps at the threshold, the sum of both terms is
continuous. This is reminiscent of the BBS contributions in the Ising model
with a boundary magnetic field~\cite{SE07}, where a cancellation of
singularities yields a smooth spectral function at the threshold.

The LDOS in the presence of a BBS is shown in Fig.~\ref{fig:plot6}. We clearly
observe a contribution within the gap $E<2\Delta$. In contrast to BBS's in the
CDW state~\cite{SEJF08,SEJF11} there is, however, no singularity at
$E=E_\mathrm{bbs}$ as the underlying processes involve the creation of a
spinon, an antisoliton, and the BBS. As before the LDOS is dominated by a
singularity at $q=0$ and we observe dispersing features following
\eqref{eq:DRE1c} and \eqref{eq:DRE1cs} respectively.

\section{Comparison to half-filled Mott insulator}\label{sec:HFMI}
The LDOS of a half-filled Mott insulator in the presence of a boundary was analyzed
in detail~\cite{endnote5} in Ref.~\onlinecite{SEJF11}. As already mentioned, the 
effective low-energy 
Hamiltonian is also given by \eqref{eq:hamiltonian}--\eqref{eq:spinhamiltonian} but the
expressions for the right- and left-moving Fermi fields in terms of the Bose fields 
differ from \eqref{eq:bosonizationR} 
and \eqref{eq:bosonizationL}. From the physical point of view right- and left-moving Fermi 
fields create and annihilate individual antisolitons in the half-filled case whereas they create 
and annihilate antisoliton pairs in the quarter-filled system. Thus the gap in the LDOS is
given by $\Delta$ in the former and $2\Delta$ in the latter system. Apart from this we find 
the following similarities and differences.

In both cases the LDOS possesses a singularity at $q=0$ which behaves as 
$N(E,2k_\mathrm{F}+q)\sim 1/\sqrt{v_\mathrm{s}q}$ independently~\cite{endnote6} 
of the Luttinger parameter $\beta$. This singularity is due to the pinning of the 
$2k_\mathrm{F}$-SDW at the boundary, which possesses a wave length $2a_0$ or
$4a_0$ respectively.

Beside this singularity the LDOS of the half-filled Mott insulator shows dispersing 
features at $E=\sqrt{\Delta^2+(v_\mathrm{c}q/2)^2}$, at $E=\Delta+v_\mathrm{s}q/2$, and, 
for $q>q_0$, at $E=\Delta\sqrt{1-(v_\mathrm{s}/v_\mathrm{c})^2}+v_\mathrm{s}q/2$. An 
interpretation is obtained by noting that the electron decomposes into one spinon and
one antisoliton, which then propagate through the system. As the leading processes involve
only individual antisolitons, the "quasi-fermionic" behavior of the antisolitons discussed at the 
end of Sec.~\ref{sec:LDOS2} does not affect the LDOS. In
contrast, in the quarter-filled system we observe only two dispersing modes following 
\eqref{eq:DRE1c} and \eqref{eq:DRE1cs} respectively (see Fig.~\ref{fig:plot2}). All other
dispersing features discussed above are strongly suppressed due to the "quasi-fermionic"
behavior of the antisolitons. 

Finally, in the presence of a BBS with energy $E_\mathrm{bbs}<\Delta$ the LDOS of the 
half-filled Mott insulator possesses a non-dispersing singularity with 
$N(E,2k_\mathrm{F}+q)\sim 1/\sqrt{E-E_\mathrm{bbs}}$ which originates from the creation of 
a spinon and the BBS.  In contrast, in the quarter-filled case the LDOS is smooth at the threshold  
$E=\Delta+E_\mathrm{bbs}$ which originates from the simultaneous creation of one 
antisoliton and the BBS.

\section{Discussion and experimental signatures}\label{sec:discussion}
In this section we compare qualitative aspects of the LDOS of various one-dimensional systems 
in the presence of a boundary or strong impurity potential. 
We discuss the spatial Fourier transform $N(E,Q)$ defined in 
\eqref{eq:defNp}. Specifically we consider
Luttinger liquids~\cite{Kivelson-03}, half-filled Mott insulators (or CDW 
states)~\cite{SEJF08,SEJF11}, and the quarter-filled Mott insulator investigated in 
Sec.~\ref{sec:LDOS}. We focus on the $Q\approx 2k_\mathrm{F}$-component of the 
LDOS as it vanishes in the translationally invariant systems and thus provides a particularly 
clean way to investigate boundary effects. (We note that the LDOS in the small-momentum 
regime $Q\approx 0$ behaves qualitatively similar.)
The first feature, which all three systems have in common, is a singularity
of the LDOS $N(E,2k_\mathrm{F}+q)$ at $q=0$, which originates in the pinning of a charge or
spin density wave at the impurity. In the case of a Luttinger liquid the strength 
of this singularity depends on the Luttinger parameter $K_\mathrm{c}$ and thus on 
the interactions between the electrons, whereas in both the half-filled
and quarter-filled Mott insulator the LDOS behaves as~\cite{endnote6} 
$N(E,2k_\mathrm{F}+q)\sim 1/\sqrt{q}$.
Furthermore, as was pointed out in Ref.~\onlinecite{EsslerKonik05}, the gap of the LDOS in 
the half-filled Mott insulator is given by the soliton mass $\Delta$, which equals the thermal
activation gap and is half of the gap observed in optical measurements. In contrast, in the 
quarter-filled system the LDOS possesses a gap of $2\Delta$, which is equal to the optical gap 
but twice the thermal gap.

\begin{figure}[tb]
\centering
\includegraphics[width=85mm,clip=true]{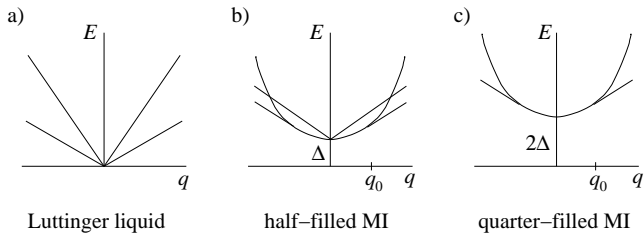}
\caption{Schematic picture of the dispersing features observable in the spatial Fourier
	transform $N(E,2k_\mathrm{F}+q)$ of the LDOS in different one-dimensional systems. 
	a) In a Luttinger liquid one finds two linearly dispersing modes. b) In a half-filled 
	Mott insulator (MI) there exists a gap $\Delta$. One observes one linear and one curved
	dispersion. From the latter a third feature splits off at $q=q_0$. c) The quarter-filled Mott
	insulator possesses a gap $2\Delta$. One further observes one curved dispersion from
	which a linear mode splits off at $q=q_0$. In both (b) and (c) the thermal activation gap 
	is given by $\Delta$.}
\label{fig:plot7}
\end{figure}
In addition, the LDOS possesses various dispersing features (see Fig.~\ref{fig:plot7}). 
In the Luttinger liquid case one observes two linearly dispersing modes corresponding
to individually propagating spin and charge degrees of freedom respectively 
(see Fig.~\ref{fig:plot7}.a). In the half-filled Mott insulator an electron decomposes into
massless spin excitations and one charge excitation with mass $\Delta$. Propagating 
spin excitations give rise to a linearly dispersing mode with velocity $v_\mathrm{s}$, while
the propagating charge excitation results in a curved dispersion relation. In addition, there 
exists a critical momentum $q=q_0$ at which the group velocity of the charge excitation
equals $v_\mathrm{s}$. At this momentum a third linearly dispersing feature splits away from 
the charge mode, which has its origin in the collective propagation of spin and charge 
degrees of freedom with equal velocities (see Fig.~\ref{fig:plot7}.b). We note that the
qualitative behavior discussed for the half-filled Mott insulator is also  expected in other systems 
possessing gapless excitations in one sector and gapped ones in the other like, for instance,
CDW states. Finally, in the quarter-filled
Mott insulator an electron decomposes into massless spin excitations and two charge excitations
with masses $\Delta$. This results in a curved dispersion originating in the propagation of 
one charge excitation while the second one stays at the position of the STM tip. This dispersion
splits at the critical momentum $q_0$ due to the collective propagation of one charge excitation
and additional spin degrees of freedom (see Fig.~\ref{fig:plot7}.c). We note that one does not 
observe individually propagating spin excitations. The relevant processes would require the
two charge excitations to possess equal momenta, which is forbidden by their "quasi-fermionic"
behavior (see Sec.~\ref{sec:LDOS2}).

To sum up, at sufficiently large momenta $q$ 
we observe two linearly dispersing modes in a Luttinger liquid, two linear and one curved 
dispersion relation in a half-filled Mott insulator, and one linear and one curved dispersion
in a quarter-filled Mott insulator. This clearly demonstrates 
that characteristic properties of the bulk state 
of matter and its electronic excitations can be infered from studying spatial 
modulations of the LDOS in the presence of an impurity. 

The LDOS discussed in this article is directly related to the tunneling current measured 
in STM experiments. If the density of states in the STM tip is approximately independent of
energy, the tunneling conductance will be given by the thermally smeared LDOS of the 
sample at the position of the tip~\cite{Fischer-07}
\begin{equation}
\frac{dI(V,x)}{dV}\propto\int dE\,f'(E-eV)\,N(E,x),
\label{eq:STM}
\end{equation}
where $f(E)$ denotes the Fermi function. $N(E,Q)$ can then easily be obtained via spatial
Fourier transformation. In this context the models discussed above apply 
to quasi-one-dimensional materials at energies above the cross-over scale to three-dimensional
behavior. This situation might be experimentally realized for example in stripe phases of 
high-temperature superconductors~\cite{Fischer-07,Kivelson-03}, 
carbon nanotubes~\cite{Lee-04,Odom-02}, Bechgaard and Fabre salts~\cite{Seo-04}, 
self-organized atomic gold chains~\cite{Schaefer-08},
and two-leg ladder materials~\cite{2leg}.

As discussed above already the knowledge of qualitative properties such as the presence 
of a spectral gap and its relation to the thermal activation gap as well as the number and 
positions of singularities can reveal substantial information about the 
electronic properties and excitations of quasi-one-dimensional materials. A more detailed 
analysis of quantitative features like the strength of singularities or the precise positions of
dispersing modes can even lead to estimates for the Luttinger parameter and the velocities
of charge and spin excitations. There are, of course, experimental limitations. Obviously
the experimental resolution and thermal broadenings have to
be sufficiently small to clearly distinguish the different singularities and dispersing modes.
Moreover, the simplyfied relation \eqref{eq:STM} does not contain~\cite{Fischer-07}
the matrix elements for the tunneling from the sample into the tip, which may introduce 
a non-trivial spatial dependence of the tunneling conductance on the LDOS and thus 
hamper a natural interpretation in terms of propagating excitations.

\section{Conclusions}
In conclusion, we have studied a quarter-filled Mott insulator in the presence
of an impurity, which was modeled by a boundary condition for the electron
fields. In this system we calculated the $2k_\mathrm{F}$-component of the
spatial Fourier transform of the LDOS using the boundary state formalism
together with a form-factor expansion. The LDOS is dominated by a singularity
at $Q=2k_\mathrm{F}$, which is indicative of a pinning of a SDW at the
impurity. Furthermore, we observed several dispersing features above the
two-soliton gap $E<2\Delta$. We identified their physical origin as spin and
charge degrees of freedom, which propagate between the STM tip and the
impurity, and attributed their relative strength to the ``quasi-fermionic''
behavior of charge excitations with equal momenta.  This also explains the
absence of dispersing features in the spectral
function~\cite{EsslerTsvelik02prl} of the translationally invariant system and
underlines that the spectral function and the LDOS probe rather different
physical processes.  We further showed that a BBS in the charge sector leads to
a non-vanishing LDOS within the two-soliton gap. 
Finally, we discussed our results in the context of other one-dimensional 
systems, i.e. Luttinger liquids and half-filled Mott insulators, and argued that 
the measurement of the spatial modulations in the LDOS can be used to 
extract detailed informations about the electronic states of quasi-one-dimensional
systems.

\section*{Acknowledgments}
I would like to thank Fabian Essler, Markus Garst, Markus Morgenstern, Alexei Tsvelik,
and particularly Volker Meden for valuable discussions and comments. This work was 
supported by the German Research Foundation (DFG) through the Emmy Noether Program.

\appendix
\section{Calculation of the Green function}\label{app:GF}
In this appendix we derive the leading terms $\mathcal{G}_k$ in the
form-factor expansion of the charge part of the Green function
\eqref{eq:GRL}, i.e. we determine the correlation function
\begin{equation}
  \Big\langle O_R(\tau,x_1)\,O_L^\dagger(0,x_2)\Big\rangle
  \label{eq:A1}
\end{equation}
in the sine-Gordon model \eqref{eq:chargehamiltonian} on the half line.  Here
the operators $O_R^\dagger$ and $O_L^\dagger$ are given by the charge parts of
\eqref{eq:bosonizationR} and \eqref{eq:bosonizationL} respectively.  We
calculate \eqref{eq:A1} using the boundary state
formalism~\cite{GhoshalZamolodchikov94} together with a form-factor
expansion~\cite{Smirnov92book,Lukyanov95,EsslerKonik05}.  The same technique
was previously applied~\cite{SEJF08,SEJF11} to determine the Green function of
a CDW state in the presence of an impurity.  Here we restrict
ourselves to the main steps of the derivation and concentrate on the
differences as compared to Ref.~\onlinecite{SEJF11}, to which we refer
implicitly for all details and definitions not discussed here.

By performing a rotation in Euclidean space we
map~\cite{GhoshalZamolodchikov94} the problem onto the sine-Gordon model on
the real axis. The boundary condition is translated into an initial condition
encoded in a boundary state. Expanding the latter in powers of the boundary
reflection matrix and inserting a resolution of the identity between the
operators in \eqref{eq:A1} we find
\begin{equation}
\label{eq:CFexp}
\Big\langle O_R(\tau,x_1)\,O_L^\dagger(0,x_2)\Big\rangle
=\sum_{n=0}^\infty\sum_{m=0}^\infty C_{n\,2m}(\tau,x_1,x_2),
\end{equation}
where we have defined the auxiliary functions
\begin{widetext}
\begin{equation}
\begin{split}
C_{n\,2m}(\tau,x_1,x_2)=&\frac{1}{2^m}\frac{1}{m!}\frac{1}{n!}
\int_{-\infty}^\infty\frac{d\theta'_1\ldots d\theta'_m}{(2\pi)^m}
\int_{-\infty}^\infty\frac{d\theta_1\ldots d\theta_{n}}{(2\pi)^{n}}\,
K^{a_1b_1}(\theta'_1)\ldots K^{a_mb_m}(\theta'_m)\label{eq:corrmn}\\*[2mm]
&\hspace{-20mm}\times
\bra{0}O_R(\tau,x_1)
\ket{\theta_{n},\ldots,\theta_{1}}_{c_n,\ldots,c_1}\;
^{c_1,\ldots,c_n}\!\bra{\theta_{1},\ldots,\theta_{n}}
O_L^\dagger(0,x_2)
\ket{-\theta'_1,\theta'_1,\ldots,-\theta'_m,\theta'_m}
_{a_1,b_1,\ldots,a_m,b_m}.
\end{split}
\end{equation}
\end{widetext}
Here the states $\ket{\theta_{n},\ldots,\theta_{1}}_{c_n,\ldots,c_1}$
(and so on) are scattering states of solitons and antisolitons, the
indices $a_i,b_i,c_i$ take the values $\mp 1$ and label the
corresponding U(1) charges, and energy and momentum of solitons and
antisolitons are given in terms of the rapidities $\theta_i$ and
$\theta'_i$ by \eqref{eq:EP}.  We note that \eqref{eq:CFexp}
constitutes an expansion in two parameters: (i) the number of
particles (with rapidities $\theta_i$) in the intermediate state and
(ii) the number of reflections of particles at the boundary [the
number of boundary reflection matrices $K(\theta'_i$)]. 

The scattering of solitons or antisolitons with rapidities $\theta_1$
and $\theta_2$ is encoded in the scattering
matrix~\cite{ZamolodchikovZamolodchikov79}
$\ket{\theta_1,\theta_2}_{ab}=S_{ab}^{cd}(\theta_1-\theta_2)
\ket{\theta_2,\theta_1}_{dc}$.  The boundary reflection
matrix~\cite{GhoshalZamolodchikov94} $K^{ab}(\theta)$ similarly
describes the reflection at the boundary,
$\ket{\theta}_a=K^{\bar{a}b}(\ii\tfrac{\pi}{2}-\theta)\ket{-\theta}_b$,
where $\bar{a}=\pm$ for $a=\mp$. For Dirichlet boundary conditions one
has~\cite{GhoshalZamolodchikov94,MattssonDorey00}
$K^{++}(\theta)=K^{--}(\theta)=0$.

Furthermore, $O(\tau,x)=e^{-xH}e^{-\ii P\tau}Oe^{\ii P\tau}e^{xH}$, where $H$
and $P$ are the Hamiltonian and the total momentum of the sine-Gordon model on
the infinite line. The operator $O_R$ ($O_L^\dagger$) creates (annihilates)
two antisolitons, i.e. increases (decreases) the U(1) charge by 2. The
necessary form factors were derived by Lukyanov and
Zamolodchikov~\cite{LukyanovZamolodchikov01}. In our conventions they are
given by
\begin{equation}
  \begin{split}
    &\bra{0}O_{R/L}\ket{\theta_1,\theta_2}_{--}\\
    &\qquad=\sqrt{Z_2}\,e^{\pm\ii\pi/8}\,
    e^{\pm(\theta_1+\theta_2)/8}\,G(\theta_1-\theta_2),
  \end{split}
  \label{eq:ff}
\end{equation}
where the normalization constant $Z_2$ and the auxiliary function $G(\theta)$
are stated in App.~\ref{app:ff}. The form factors satisfy the form-factor
axioms as stated in Ref.~\onlinecite{SEJF11} with Lorentz spin $s(O_{R/L})=\pm
1/4$ and semi-locality factor $l_-(O_{R/L})=e^{\pm\ii\pi/4}$.  In the
following we will evaluate the terms $C_{20}$, $C_{22}$, and $C_{24}$ which
result in $\mathcal{G}_0$, $\mathcal{G}_1$, and $\mathcal{G}_2$, respectively.

\subsection{Derivation of $\boldsymbol{\mathcal{G}_0}$}
The first term to be evaluated is $C_{20}$. It does not contain any
information about the boundary in the charge sector, i.e. it contains no
boundary reflection matrices. A similar term was calculated in
Ref.~\onlinecite{EsslerTsvelik02prl} in the derivation of the spectral
function in the translationally invariant system. After shifting the contour of
integration, $\theta_{1,2}\rightarrow\theta_{1,2}+\ii\pi/2$, we directly
obtain
\begin{equation*}
  \begin{split}
  \mathcal{G}_0=C_{20}=\frac{Z_2}{2}e^{\ii\pi/4}\,&
  \int_{-\infty}^\infty\frac{d\theta_1d\theta_2}{(2\pi)^{2}}\,
  \big|G(\theta_1-\theta_2)\big|^2\\
  &\;\times e^{\ii\frac{\Delta}{v_\mathrm{c}}r\sum_i\sinh\theta_i}\,
  e^{-\Delta\tau\sum_i\cosh\theta_i},
  \end{split}
\end{equation*}
where the center-of-mass coordinates are defined by $x=(x_1+x_2)/2<0$ and
$r=x_1-x_2<0$.  Analytical continuation $\tau\rightarrow\ii t$ to real times,
taking the limit $r\rightarrow 0$, and Fourier transformation \eqref{eq:defNp}
with respect to $t$ and $x$ yield the first term $N_0(E,2k_\mathrm{F}+q)$ in
the form-factor expansion of the LDOS \eqref{eq:LDOSresult}.

\subsection{Derivation of $\boldsymbol{\mathcal{G}_1}$}
We start with $C_{22}$ defined in \eqref{eq:corrmn}. As the operator $O_R$
creates two antisolitons we deduce $c_1=c_2=-$;
$K^{++}(\theta')=K^{--}(\theta')=0$ yields $b=\bar{a}$. The matrix element of
$O_L^\dagger$ contains incoming and outgoing particles and thus possesses
kinematical poles. We deal with these terms following
Smirnov~\cite{Smirnov92book} and analytically continue the form factor as (see
Ref.~\onlinecite{SEJF11} for a detailed discussion of this procedure)
\begin{eqnarray}
  & &\hspace{-10mm}
  ^{--}\!\bra{\theta_1,\theta_2}O_L^\dagger\ket{-\theta',\theta'}_{a\bar{a}}
  \nonumber\\*
  &=&^{--}\!\bra{\theta_1\!+\!\ii 0,\theta_2\!+\!\ii 0}O_L^\dagger
  \ket{-\theta',\theta'}_{a\bar{a}}\nonumber\\*
  & &+2\pi\,\delta(\theta_1-\theta')\,\delta_{-\bar{a}}\;
  ^-\!\bra{\theta_2}O_L^\dagger\ket{-\theta'}_a\label{eq:regG1}\\*
  & &+2\pi\,\delta(\theta_1+\theta')\,
  \delta_{-b}\,S_{a\bar{a}}^{bc}(-2\theta')\;
  ^-\!\bra{\theta_2}O_L^\dagger\ket{\theta'}_c\nonumber\\*
  & &+2\pi\,\delta(\theta_2-\theta')\,\delta_{e\bar{a}}\,
  S_{--}^{de}(\theta_1-\theta_2)\;
  ^d\!\bra{\theta_1}O_L^\dagger\ket{-\theta'}_a\nonumber\\*
  & &+2\pi\,\delta(\theta_2+\theta')\,\delta_{be}\,
  S_{a\bar{a}}^{bc}(-2\theta')\nonumber\\*
  & &\qquad\times S_{--}^{de}(\theta_1-\theta_2)\;
  ^d\!\bra{\theta_1}O_L^\dagger\ket{\theta'}_c.\nonumber
\end{eqnarray}
Here we have already omitted the positive imaginary parts of the rapidities in
the form factors in the 2nd to 5th lines as they do not possess kinematical
poles.  Inserting \eqref{eq:regG1} into $C_{22}$ yields to different
contributions: The first line leads to a term containing three integrations
over rapidities and is thus a sub-leading correction; we will not evaluate it
here.  The 2nd to 5th lines yield using the boundary cross-unitarity
condition~\cite{GhoshalZamolodchikov94}
$K^{ab}(\theta)=S^{ab}_{cd}(2\theta)\,K^{dc}(-\theta)$ as well as
unitarity~\cite{ZamolodchikovZamolodchikov79}
$S_{a_1a_2}^{c_1c_2}(\theta)S_{c_1c_2}^{b_1b_2}(-\theta)=
\delta_{a_1}^{b_1}\delta_{a_2}^{b_2}$
\begin{equation}
  \begin{split}
  &\int_{-\infty}^\infty\frac{d\theta'}{2\pi}\frac{d\theta}{2\pi}\,
  K^{+-}(\theta')\,\bra{0}O_R\ket{\theta,\theta'}_{--}
  \,^-\!\bra{\theta}O_L^\dagger\ket{-\theta'}_+\\
  &\qquad\times e^{\ii\Delta\tau(\sinh\theta+\sinh\theta')}\,
  e^{\frac{\Delta}{v_\mathrm{c}}(r\cosh\theta+2x\cosh\theta')}.
  \label{eq:G2A5}
  \end{split}
\end{equation}
In order to proceed let us first assume $\Phi_\mathrm{c}^0=0$. This implies
that the boundary reflection matrix does not depend on the U(1) charge, i.e.
$K^{+-}(\theta')=K^{-+}(\theta')=K(\theta')$, where $K(\theta)$ is explicitly
stated in App.~\ref{app:ff}. In particular, $K(\theta')$ is analytic in the
physical strip $0\le\mathrm{Im}\,\theta'\le\pi/2$. Now we shift the contours
of integration $\theta,\theta'\rightarrow\theta,\theta'+\ii\pi/2$ (we note
that the contribution originating in the first term of \eqref{eq:regG1} does
not possess any poles in the physical strip
$0\le\mathrm{Im}\,\theta_i\le\pi/2$) and apply the crossing relation in the
second form factor,
\begin{equation*}
  \begin{split}
    &^-\!\bra{\theta+\ii\tfrac{\pi}{2}}
    O_L^\dagger\ket{-\theta'-\ii\tfrac{\pi}{2}}_+
    =^+\!\bra{-\theta'+\ii\tfrac{\pi}{2}}O_L\ket{\theta-\ii\tfrac{\pi}{2}}_-^*\\
    &\qquad=e^{-\ii\pi/4}\,\bra{0}O_L
    \ket{-\theta'+\ii\tfrac{3\pi}{2},\theta-\ii\tfrac{\pi}{2}}_{--}^*,
  \end{split}
\end{equation*}
where the phase factor $e^{-\ii\pi/4}$ is due to the
semi-locality~\cite{Smirnov92book,Lukyanov95,EsslerKonik05} of the
operator $O_L$ with respect to the fundamental fields creating the excitations.
Finally we use \eqref{eq:ff} to obtain
\begin{equation*}
  \begin{split}
  \mathcal{G}_1=&\,Z_2\,e^{\ii\pi/4}
  \int_{-\infty}^\infty\frac{d\theta'}{2\pi}\frac{d\theta}{2\pi}\,
  K(\theta'+\ii\tfrac{\pi}{2})\,G(\theta-\theta')\,G(\theta+\theta')^*\\
  &\qquad\times e^{\theta'/4}\,
  e^{\ii\frac{\Delta}{v_\mathrm{c}}(r\sinh\theta+2x\sinh\theta')}\,
  e^{-\Delta\tau(\cosh\theta+\cosh\theta')}.
  \end{split}
\end{equation*}
Fourier transformation \eqref{eq:defNp} directly yields the second term 
$N_1(E,2k_\mathrm{F}+q)$ of the LDOS.

On the other hand, if $\Phi_\mathrm{c}^0<-\beta\pi$, the boundary reflection
matrix $K^{+-}(\theta')$ possesses a pole at $\theta'=\ii\gamma$. At the LEP
the residue is given by $2\ii\cos(\Phi_\mathrm{c}^0/\sqrt{2})$ (see
\eqref{eq:KLEP}). Thus when shifting the contours of integration in
\eqref{eq:G2A5} we obtain the additional BBS contribution
\begin{equation*}
  \begin{split}
    \mathcal{G}_3=&-2\ii\,Z_2\cos\frac{\Phi_\mathrm{c}^0}{\sqrt{2}}\,
    e^{-\ii(\pi-\sqrt{2}\Phi_\mathrm{c}^0)/8}
    \,e^{-E_\mathrm{bbs}\tau}\,e^{\kappa_\mathrm{bbs}x}\\
    &\hspace{-8mm}\times\!\int_{-\infty}^\infty\frac{d\theta}{2\pi}\,
    \biggl|
    \sinh\frac{\theta\!+\!\frac{\ii}{2}(\pi\!+\!\sqrt{2}\Phi_\mathrm{c}^0)}{2}
    \biggr|^2\,
    e^{\ii\frac{\Delta}{v_\mathrm{c}}r\sinh\theta}\,
    e^{-\Delta\tau\cosh\theta}.
  \end{split}
\end{equation*}
Fourier transformation yields $N_3(E,2k_\mathrm{F}+q)$.

\subsection{Derivation of $\boldsymbol{\mathcal{G}_2}$}
Finally we calculate the leading term involving the reflection of two
antisolitons, which originates from $C_{24}$. The analytic continuation of the
form factor of $O_L^\dagger$ reads
\begin{eqnarray}
  & &\hspace{-7mm}
  ^{--}\!\bra{\theta_1,\theta_2}O_L^\dagger
  \ket{-\theta'_1,\theta'_1,-\theta'_2,\theta'_2}_{a\bar{a}b\bar{b}}
  \nonumber\\*
  &=&^{--}\!\bra{\theta_1\!+\!\ii 0,\theta_2\!+\!\ii 0}O_L^\dagger
  \ket{-\theta'_1,\theta'_1,-\theta'_2,\theta'_2}_{a\bar{a}b\bar{b}}
  \nonumber\\*
  & &+S_{-+}^{-+}(\theta'_1+\theta'_2)\,\delta_{+a}\delta_{+b}
  \label{eq:regG2}\\*
  & &\qquad\times
  ^{--}\!\bra{\theta_1,\theta_2}\theta'_1,\theta'_2\rangle_{--}\;
  \bra{0}O_L^\dagger\ket{-\theta'_1,-\theta'_2}_{++}\nonumber\\*
  & &+S_{a\bar{a}}^{-+}(-2\theta'_1)\,S_{-+}^{-+}(-\theta'_1+\theta'_2)
  \,\delta_{+b}\nonumber\\*
  & &\qquad\times
  ^{--}\!\bra{\theta_1,\theta_2}-\theta'_1,\theta'_2\rangle_{--}\;
  \bra{0}O_L^\dagger\ket{\theta'_1,-\theta'_2}_{++}\nonumber\\*
  & &+S_{b\bar{b}}^{-+}(-2\theta'_2)\,S_{-+}^{-+}(\theta'_1-\theta'_2)
  \,\delta_{+a}\nonumber\\*
  & &\qquad\times
  ^{--}\!\bra{\theta_1,\theta_2}\theta'_1,-\theta'_2\rangle_{--}\;
  \bra{0}O_L^\dagger\ket{-\theta'_1,\theta'_2}_{++}\nonumber\\*
  & &+S_{a\bar{a}}^{-+}(-2\theta'_1)\,S_{b\bar{b}}^{-+}(-2\theta'_2)\,
  S_{-+}^{-+}(-\theta'_1-\theta'_2)\nonumber\\*
  & &\qquad\times
  ^{--}\!\bra{\theta_1,\theta_2}-\theta'_1,-\theta'_2\rangle_{--}\;
  \bra{0}O_L^\dagger\ket{\theta'_1,\theta'_2}_{++},\nonumber
\end{eqnarray}
where the scalar products are evaluated using the Faddeev-Zamolodchikov
algebra~\cite{ZamolodchikovZamolodchikov79,Faddeev80}
\begin{equation*}
  \begin{split}
  ^{--}\!\bra{\theta_1,\theta_2}\theta'_1,\theta'_2\rangle_{--}=&\,
  (2\pi)^2\bigl[\delta(\theta_1-\theta'_2)\,\delta(\theta_2-\theta'_1)\\
  &\hspace{-15mm}
  +S_{--}^{--}(\theta'_1-\theta_2)\,\delta(\theta_1-\theta'_1)\,
  \delta(\theta_2-\theta'_2)\bigr].  
  \end{split}
\end{equation*}
Therefore the 2nd to 5th line in \eqref{eq:regG2} yield (we also use the
scattering axiom for the first form factor)
\begin{equation}
  \begin{split}
    &\frac{1}{2}\int_{-\infty}^\infty\frac{d\theta'_1d\theta'_2}{(2\pi)^2}\,
    K^{+-}(\theta'_1)\,K^{+-}(\theta'_2)\,S_{-+}^{-+}(\theta'_1+\theta'_2)\\
    &\quad\times
    \bra{0}O_R\ket{\theta'_1,\theta'_2}_{--}\,\bra{0}O_L^\dagger\ket{-\theta'_1,-\theta'_2}_{++}\\
    &\quad\times
    e^{\ii\Delta\tau\sum_i\sinh\theta'_i}\,
    e^{2\frac{\Delta}{v_\mathrm{c}}x\sum_i\cosh\theta'_i}.
    \label{eq:G22}
  \end{split}
\end{equation}
Let us again first assume $\Phi_\mathrm{c}^0=0$. If we shift the contours of
integration $\theta'_i\rightarrow\theta'_i+\ii\pi/2$ and use the
crossing relation as well as a Lorentz transformation,
\begin{equation*}
  \begin{split}
  &\bra{0}O_L^\dagger
  \ket{-\theta'_1-\ii\tfrac{\pi}{2},-\theta'_2-\ii\tfrac{\pi}{2}}_{++}\\
  &\qquad\qquad=
  e^{-\ii\pi/8}\,\bra{0}O_L\ket{-\theta'_2,-\theta'_1}_{--}^*,
  \end{split}
\end{equation*}
we obtain
\begin{equation*}
  \begin{split}
    \mathcal{G}_2=&\,
    \frac{Z_2}{2}\,e^{\ii\pi/4}
    \int_{-\infty}^\infty\frac{d\theta'_1d\theta'_2}{(2\pi)^2}\,
    K(\theta'_1+\ii\tfrac{\pi}{2})\,K(\theta'_2+\ii\tfrac{\pi}{2})\\
    &\!\!\!\!\times S_{-+}^{-+}(\theta'_1+\theta'_2+\ii\pi)\,
    \big|G(\theta'_1-\theta'_2)\big|^2\\
    &\!\!\!\!\times e^{(\theta'_1+\theta'_2)/4}\,
    e^{2\ii\frac{\Delta}{v_\mathrm{c}}x\sum_i\sinh\theta'_i}\,
    e^{-\Delta\tau\sum_i\cosh\theta'_i}.
  \end{split}
\end{equation*}
$N_2(E,2k_\mathrm{F}+q)$ is now readily obtained via Fourier transformation.

Starting from \eqref{eq:G22} the second term $N_4(E,2k_\mathrm{F}+q)$ related
to the BBS is obtained by picking up the poles of $K^{+-}(\theta'_1)$ and
$K^{+-}(\theta'_2)$ at $\theta'_{1,2}=\ii\gamma$ respectively. At the LEP the
result is
\begin{equation*}
  \begin{split}
    \mathcal{G}_4=&\,2\ii\,Z_2\cos\frac{\Phi_\mathrm{c}^0}{\sqrt{2}}\,
    e^{-\ii(\pi-\sqrt{2}\Phi_\mathrm{c}^0)/8}
    \,e^{-E_\mathrm{bbs}\tau}\,e^{\kappa_\mathrm{bbs}x}\\
    &\times\int_{-\infty}^\infty\frac{d\theta'}{2\pi}\,
    K^{+-}(\theta'+\ii\tfrac{\pi}{2})\,
    \sinh^2\frac{\theta'-\frac{\ii}{2}(\pi+\sqrt{2}\Phi_\mathrm{c}^0)}{2}\\
    &\qquad\qquad\times
    e^{\theta'/4}\,e^{\ii\frac{\Delta}{v_\mathrm{c}}r\sinh\theta'}\,
    e^{-\Delta\tau\cosh\theta'}.
  \end{split}
\end{equation*}

\pagebreak
\section{Explicit expressions for form factors and scattering
  matrices}\label{app:ff}
In this appendix we collect some formulas on the sine-Gordon theory used in
the present article. The scattering matrix between solitons and antisolitons
possesses the integral representation~\cite{EsslerKonik05}
\begin{equation*}
  S_0(\theta)=-\exp\left[\ii\int_0^\infty\frac{dt}{t}\,
    \sin\frac{\theta t}{\pi\xi}\,
    \frac{\sinh\bigl(\frac{t}{2}(1-\frac{1}{\xi})\bigr)}
  {\sinh\frac{t}{2}\,\cosh\frac{t}{2\xi}}\right],
\end{equation*}
where $\xi=\beta^2/(1-\beta^2)$.  A similar integral representation for the
boundary reflection matrix reads~\cite{Caux-03} (for Dirichlet boundary
conditions with $\Phi^0_\mathrm{c}=0$)
\begin{eqnarray*}
  K(\theta)\!\!\!&=&\!\!\!
  -\cos\left(\frac{\pi}{2\xi}+\ii\frac{\theta}{\xi}\right)
  R_0\bigl(\ii\tfrac{\pi}{2}-\theta\bigr)\,
  \sigma\bigl(\ii\tfrac{\pi}{2}-\theta\bigr),\\
  R_0(\theta)\!\!\!&=&\!\!\!\exp\!\left[2\ii\int_0^\infty\frac{dt}{t}\,
    \sin\frac{\theta t}{\pi\xi}\,\sinh\frac{3t}{4\xi}\,
    \frac{\sinh\bigl(\frac{t}{4}(1-\frac{1}{\xi})\bigr)}
    {\sinh\frac{t}{4}\,\sinh\frac{t}{\xi}}\right],\\
  \sigma(\theta)\!\!\!&=&\!\!\!\exp\!\left[2\ii\int_0^\infty\frac{dt}{t}\,
    \sin\frac{\theta t}{\pi\xi}\,
    \frac{\sinh\bigl(\frac{t}{\xi}(1+\ii\frac{\theta}{\pi})\bigr)}
    {\sinh t\,\cosh\frac{t}{\xi}}\right].
\end{eqnarray*}
The auxiliary function $G(\theta)$ appearing in the form factors \eqref{eq:ff}
is given by~\cite{LukyanovZamolodchikov01}
\begin{eqnarray*}
  G(\theta)&=&-\ii\,\mathcal{C}_1\sinh\frac{\theta}{2}\\*
  & &\hspace{-10mm}\times\exp\left[\int_0^\infty\frac{dt}{t}\,
    \frac{\sinh^2\bigl(t(1+\ii\frac{\theta}{\pi})\bigr)\,
    \sinh\bigl(t(\xi-1)\bigr)}
    {\sinh(2t)\,\cosh t\,\sinh(t\xi)}\right],\\
  \mathcal{C}_1&=&\exp\left[-\int_0^\infty\frac{dt}{t}\,
    \frac{\sinh^2\frac{t}{2}\,\sinh\bigl(t(\xi-1)\bigr)}
    {\sinh(2t)\,\cosh t\,\sinh(t\xi)}\right],
\end{eqnarray*}
while for the normalization factor $Z_2$ one finds numerically from the integral repesentation
derived in Ref.~\onlinecite{LukyanovZamolodchikov01}
\begin{eqnarray*}
  &&Z_2\approx 7.7320\,\Delta^{1681/1440}\quad\text{for}\;\beta^2=0.9,\\
  &&Z_2\approx 20.1857\,\Delta^{1649/1120}\quad\text{for}\;\beta^2=0.7.
\end{eqnarray*}
The expressions at the LEP are obtained by setting $\xi=1$ ($\beta^2=1/2$),
i.e. $S_0(\theta)=-1$, $K(\theta)=\ii\tanh\frac{\theta}{2}$,
$G(\theta)=-\ii\sinh\frac{\theta}{2}$, and $Z_2\approx
38.5519\,\Delta^{65/32}$.



\begin{thebibliography}{10}
\bibitem{endnote1} 
  Starting from (4.24) in Ref.~\onlinecite{Giamarchi04} with $n=2$ we rescale
  the Bose fields according to
  $\phi_\rho\rightarrow\beta\Phi_\mathrm{c}/\sqrt{32}$ and
  $\theta_\rho\rightarrow\Theta_\mathrm{c}/\sqrt{2}\beta$ with $\beta^2=4K$.
  The double-Umklapp term is relevant for $K<1/4$, i.e. for $\beta<1$.
\bibitem{endnote2}
  The LDOS for $E<0$ can be analyzed in the same way.
\bibitem{endnote3}
  The small momentum regime is dominated by a similar singularity, $N(E,Q)\sim
  N_0(E,Q)\sim 1/Q$.
\bibitem{endnote4}
  This feature is more pronounced in $\text{Re}\,N_0(E,2k_\mathrm{F}+q)$.
\bibitem{endnote5} 
  The half-filled Mott insulator is equivalent to the attractive regime of a CDW state 
  discussed in Sec.~IV.B of Ref.~\onlinecite{SEJF11} with $k_\mathrm{F}=\pi/2a_0$ 
  and $K_\mathrm{c}=1$.
\end{thebibliography}
\end{document}